\documentclass[aip,jcp,reprint]{revtex4-1}
\usepackage{amsmath}
\usepackage{amsfonts}
\usepackage{amssymb}
\usepackage{graphics}
\usepackage{float}
\usepackage{graphicx}
\usepackage[compact]{titlesec}
\usepackage{natbib}
\usepackage{threeparttable}
\usepackage[titletoc,title]{appendix}
\usepackage{lipsum}
\usepackage{varwidth}
\usepackage{tabularx}
\usepackage{color}

\newcommand{\etal}{\textit{et al}. }

\begin{document}
\title{The electrothermal conductance and heat capacity of black phosphorus}
\author{Parijat Sengupta}
%\affiliation{Photonics Center, Boston University, Boston, MA 02215.}
\affiliation{Department of Electrical and Computer Engineering \\
University of Illinois at Chicago, Chicago, IL 60607.}
\author{Saptarshi Das}
\affiliation{Department of Engineering Science and Mechanics \\
Pennsylvania State University, State College, PA 16802.}
\author{Junxia Shi}
\email{lucyshi@uic.edu}
\affiliation{Department of Electrical Engineering \\
University of Illinois at Chicago, Chicago, IL 60607.}

\begin{abstract}
We study a thermal gradient induced current $\left(I_{th}\right)$ flow in potassium-doped two-dimensional anisotropic black phosphorus (BP) with semi-Dirac dispersion. The prototype device is a BP channel clamped between two contacts maintained at unequal temperatures. The choice of BP lies in the predicted efficient thermolectric behaviour. A temperature-induced difference in the Fermi levels of the two contacts drives the current (typified by the electro-thermal conductance) which we calculate using the Landauer transport equation. The current shows an initial rise when the device is operated at lower temperatures. The rise stalls at progressively higher temperatures and $I_{th}$ acquires a plateau-like flat profile indicating a competing effect between a larger number of transmission modes and a corresponding drop in the Fermi level difference between the contacts. The current is computed for both \textit{n}- and \textit{p}-type BP and the difference thereof is attributed to the particle-hole asymmetry. The utility of such calculations lie in conversion of the heat production and an attendant temperature gradient in miniaturized devices to useful electric power and a possible realization of solid-state Peltier cooling. Unlike the flow of $I_{th}$ that removes heat, the ability of a material to maintain a steady temperature is reflected in its heat capacity which is formulated in this work for BP via a Sommerfeld expansion. In the concluding part, we draw a microscopic connection between the two seemingly disparate processes of heat removal and absorption by pinning down their origin to the underlying density of states. Finally, a qualitative analysis of a Carnot-like efficiency of the considered thermoelectric engine is performed drawing upon the previous results on thermal current and heat capacity. 
\end{abstract}
\maketitle

\vspace{0.2cm}
\section{Introduction}
\label{sec1}
\vspace{0.2cm}
The miniaturization of circuit components introduces the problem of localized heating~\cite{ma2007heat,mcglen2004integrated} that can give rise to temperature overshoots and degrade their overall life span. Such problems have necessitated the need for improved cooling schemes~\cite{vandersande1996thermal} to hold the operating temperature to reasonable limits. The overarching goal is to drive the heat from localized high-temperature regions - more commonly known as hot-spots - to draw level with ambient conditions. While the generated heat can be simply removed, a profitable spin-off is to transform the heat current~\cite{bell2008cooling} into electric power taking advantage of the Seebeck effect~\cite{behnia2015fundamentals} which describes current flow in a looped conductor under a finite temperature gradient. This constitutes the basis for electric-thermal energy conversion.~\cite{iwanaga2011high,minnich2009bulk,zhang2015thermoelectric} The converse of this, the Peltier effect, manifests as a reduction (or gain) in temperature via pumping of heat when charge current passes through a junction of two dissimilar materials with distinct thermal behaviour. The optimization of the Seebeck and Peltier power conversion techniques have understandably received much attention, they have, however, acquired greater significance as newer materials - notably graphene~\cite{ghosh2008extremely} - hold promise of better thermoelectric operation, demonstrated by a higher figure of merit, \textit{ZT}. In addition to the two-dimensional (2D) graphene which has a large Seebeck coefficient~\cite{dragoman2007giant,yokomizo2013giant}, in part, attributable to its linearly dispersing bands, more unconventional 2D cases not fully-explored yet also exist with `hybrid' band profiles. For instance, Pardo \etal unveiled a semi-Dirac dispersion~\cite{pardo2009half,pardo2010metal} marked by a co-existence of the linear Dirac-type linear and parabolic branches in VO$_{2}$ layers embedded in TiO$_{2}$. In recent times, more evidence of such dispersion was also established in an experimental (and later through first-principles calculations) study of potassium-doped multi-layered black phosphorus (BP) by Kim \etal; they showed~\cite{baik2015emergence,kim2015observation} that the armchair (\textit{y}-axis) direction carried a linear dispersion while acquiring a parabolic character along the zigzagged \textit{x}-axis. 

\begin{figure}[t!]
\includegraphics[scale=0.85]{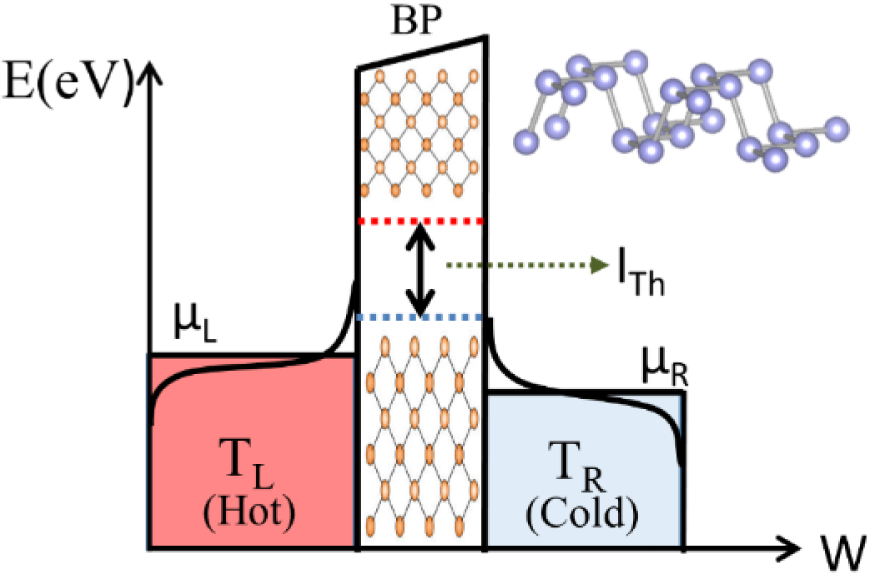}
\vspace{-0.17cm}
\caption{The schematic illustrates the suggested arrangement of a BP channel flanked between left and right contacts maintained at temperatures $ T_{L} > T_{R}$ and electrochemical potentials $ \mu_{L} $ and $ \mu_{R} $, respectively. The curved lines denote the smeared Fermi function at a finite temperature. A thermally-excited electron (hole) current $\left(I_{th}\right)$ flows from the contact at a higher (lower) temperature. The inset shows the puckered unit cell of a single layer BP. The channel dimensions (in appropriate units) are $ W $ and $ L $ along the \textit{x}- and \textit{y}-axes.}
\label{scther}
\vspace{-0.5cm}
\end{figure}

In this work, using BP as the representative material with a semi-Dirac dispersion, we quantitatively assess the electrothermal conductance and the attendant current that flows solely on account of a temperature gradient induced difference in Fermi levels. The thermally-driven current calculations are done using the Landauer equation.~\cite{datta1997electronic} In principle, this also serves as an illustration of the microscale energy `harvesting' paradigm (thermal-to-electric) via a prototypical Seebeck-like power generator (for a schematic representation, see Fig.~\ref{scther}) with applications to solid state cooling through heat removal via the Peltier process. In the second half, a related calculation estimates the heat capacity to gauge its effectiveness as a coolant; an action that is principally converse to that of heat removal via a temperature-driven current and essentially measures the material's inherent ability to absorb heat. We compute the heat capacity for BP using the well-known formula derivable from a Sommerfeld expansion.~\cite{grosso2014solid}  
As for the choice of BP~\cite{ling2015renaissance,xia2014rediscovering,liu2017temperature} for these calculations, we explain by briefly expounding on its unique thermal and electrical transport characteristics that promise improved thermoelectric behaviour. The efficiency of thermoelectric processes, in general, are marked by their \textit{ZT}, which quantifies the performance metric manifesting as an interplay between the electric and thermal conductivities. A significant value for \textit{ZT} is achievable for a high electric conductivity $\left(\sigma\right)$ and a concurrent low value of its thermal counterpart $\left(\kappa\right)$. In most materials, however, a simultaneous fulfillment of this condition is difficult to accomplish, usually prompting a trade-off to optimize the ratio, $\sigma/\kappa $. BP, in contrast, because of the intrinsic anisotropy~\cite{liu2016mobility} has a pronounced electric conductivity in the armchair direction while significant thermal transport occurs along the zigzag-axis.~\cite{fei2014enhanced} A combination of externally impressed thermal and potential gradients along the zigzag and the armchair, respectively, can therefore aid in sharply enhancing $\sigma/\kappa $ and consequently \textit{ZT}. The uptick in \textit{ZT} - predicted to touch 2.5 - offers a viable alternative to existing thermoelectric materials.~\cite{saito2016gate,lv2014large,zhang2014phosphorene} 

The calculations show that the thermal current is marked by a steady increase when the prototype device (see Fig.~\ref{scther}) is operated at lower temperatures. However, the rise tails off for elevated temperatures and a plateau-like curve is obtained. This is explained by noting that the difference in Fermi levels between the two contacts shows a slow increase at higher temperature leading to saturating current profile. We point out that the Fermi level difference serves as the greater driving force vis-\'a-vis the increased number of transmission modes at higher temperatures. Additionally, we find, as anticipated, a higher thermal current for a larger temperature difference between the contacts. The heat capacity, derived from the density of states (DOS), similarly shows an increase with energy around the Dirac crossing since at low values of momentum the linear branch dominates the quadratic contribution. The DOS increases linearly with energy for linear bands. A comparative study of the heat capacity of BP with a two-dimensional quantum well of GaAs, a prototypical III-V compound with very parabolic conduction bands around the Brillouin zone, reveals a three-order higher value for the former suggesting superior performance as a coolant. In general, a semi-Dirac system like \textit{K}-doped BP, in principle, has a higher heat capacity compared to materials identified by a purely parabolic dispersion. We also explain how the thermal current and the heat capacity, independent of the chosen material, and outwardly dissimilar in their functionality, owe their modulations to the underlying arrangement of electron states. A short qualitative theoretical discussion visualizes the arrangement as an implementation of a Carnot-like heat engine with tunable efficiency. A caveat about what follows must be made explicit here: The current flow, including heat and charge, are essentially non-equilibrium phenomena; however, in this work, we tacitly assume that all thermoelectric phenomenon is modeled for situations not far away from equilibrium such that the dispersion relation and Fermi distribution function (which is strictly defined for equilibrium situations) are considered valid for all cases described herein. 

\vspace{0.3cm}
\section{Analytic formulation}
\begin{figure}[t!]
\includegraphics[scale=0.7]{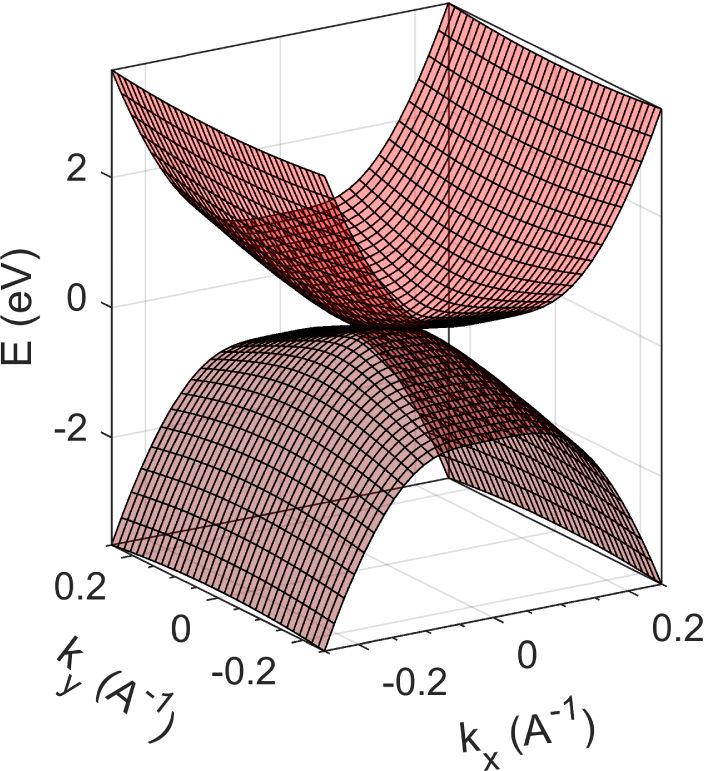}
\vspace{-0.17cm}
\caption{An illustrative representation of a semi-Dirac dispersion is shown using BP as a candidate example. Four BP layers stacked together along an out-of-plane axis (inter-layer spacing is $ 5.3\,\AA $) give rise to a semi-Dirac dispersion with a finite gap adjustable through \textit{K}-doping. The single-layer BP is actually a double layered structure and has two P–P bonds. The shorter bond (bond length is $ 2.22\,\AA $) connects the nearest P atoms in the same plane while the longer bond ($ 2.24 \AA $) connects atoms located in the top and bottom layers of the unit cell. The lower panel is the dispersion of a four-layer BP; for greater visual clarity that enhances the zigzag parabolic branch vis-\`a-vis the linear term along armchair, we artificially set the effective mass  to $ m_{e} = 0.06m_{0} $. Here, $ m_{0} $ is the free electron mass and $ v_{f} $ is $ 5.6 \times 10^{5} ms^{-1} $.}
\label{dispfig}
\vspace{-0.6cm}
\end{figure}

\vspace{0.3cm}
A semi-Dirac material~\cite{banerjee2012phenomenology} is characterized by a set of parabolic and linear bands along two mutually perpendicular directions. The minimal Hamiltonian is
\begin{equation}
\mathcal{H} = \left(\Delta + \alpha k_{x}^{2}\right)\tau_{x} + \beta k_{y}\tau_{y}.
\label{ham1}
\end{equation}
In Eq.~\ref{ham1}, the coefficient $ \alpha = \hbar^{2}/2m_{e} $ and $ \beta = \hbar v_{f} $. The Fermi velocity is $ v_{f} $, the effective mass is $ m_{e} $, and $ \tau_{x,y} $ represent the iso-spin matrices. For a generalized case, we also include a band gap $\left(\Delta\right)$ in conjunction with the parabolic set of bands. The dispersion is straightforward to obtain, we have
\begin{equation}
E\left(k\right) = \pm\,\sqrt{\left(\Delta + \alpha k_{x}^{2}\right)^{2} + \left(\beta k_{y}\right)^{2}}.
\label{bpdp}
\end{equation}
The + (-) sign in the energy expressions denote the conduction (valence) state. The dispersion by letting $ \Delta \rightarrow 0 $ in Eq.~\ref{bpdp} clearly points to mass-less Dirac Fermions along the armchair direction (\textit{y}-axis) while the zigzag axis (\textit{x}-axis) hosts the conventional `massive' parabolic electron. The band gap in Eq.~\ref{bpdp} is tunable in case of BP through \textit{K}-doping which vanishes (see Supporting Info, Ref.~\onlinecite{baik2015emergence}) at a threshold value. The above dispersion relation which holds true for a four-layered BP (adequately \textit{K}-doped such that the band gap vanishes) is plotted in Fig.~\ref{dispfig} with appropriate parameter values in the accompanying caption.

For a theoretical determination of the temperature-driven current, we must start from the dispersion relationship describing the electrons and holes. The dispersion, however, must be asymmetric (does not possess particle-hole symmetry) for a non-vanishing current.~\cite{zlatic2014modern} This constraint arises, as is well-known, because electrons and holes guided by their respective dispersion flow in opposite directions under an established electrical bias; evidently a symmetric relationship exactly cancels each contribution leading to a vanishing thermal current. For the case of gapped BP, the contribution of the electron and hole parts would be unequal and dependent on the position of the Fermi level. An \textit{n}(\textit{p})-type BP would receive significant contribution to the thermal current from the flow of electron (hole) in conduction (valence) states. However, in the eventuality that the gap diminishes to zero (by an electric field, which in our case comes from \textit{K}-doping) and the Fermi level is aligned exactly with the semi-Dirac crossing (see Fig.~\ref{dispfig}), the symmetry of the electron and hole parts would ensure a cessation of the thermal current. If the assertion that a finite current always flows regardless of the band gap and Fermi level is true everywhere, it is necessary to break the particle-hole symmetry via an external disturbance. For the most generalized case, we consider the situation where the BP layers are epitaxially grown on a ferromagnetic substrate and note that an asymmetry always exists. To see this, observe that the ferromagnetic substrate modifies the Hamiltonian through an inclusion of the additional exchange energy term, $\left(\Delta_{ex}\tau_{0}\otimes\sigma_{z}\right)$. Here, $\tau_{0} $ is the $ 2 \times 2 $ identity matrix, $ \Delta_{ex} $ is the strength of the exchange field, and $ \sigma_{z} $ is the \textit{z}-component of the Pauli spin-matrix. The revised dispersion relation is then, $ E^{s}\left(k\right) = \eta\Delta_{ex} \pm E $, which is manifestly asymmetric for electron and hole states. Here, $ \eta = \pm 1 $ identifies the spin-up/down (+1/-1) polarized dispersion curves. To distinguish from the pristine case described by Eq.~\ref{bpdp}, the superscript `s' in the energy expression indicates a modification by the ferromagnetic substrate. In writing the exchange term, it is tacitly assumed that both sub-lattices ($ A $ and $ B $) of BP are subjected to the same exchange field. As a useful digression, we  may note that asymmetry can also be introduced in BP flakes through transfer of strain via a stretched substrate, for example, polyethylene terephthalate substrates.~\cite{zhang2017plane} With this asymmetric dispersion as a basis, we attempt to compute the thermal current (see Fig.~\ref{scther}) for a film of \textit{K}-doped semi-Dirac BP juxtaposed between two contacts maintained at different temperatures. This gives us the electrothermal conductance and current behaviour.
 
\vspace{0.3cm}
\section{Thermal current}
\vspace{0.3cm}
The thermal current that flows between the two contacts is obtained using the Landauer-B\"uttiker formalism (LBF).~\cite{niu2015enhanced} For each spin component (assuming the bands are spin-split when BP is grown on a ferromagnetic substrate), it is:
\begin{equation}
I_{th} = \dfrac{e}{h}\int_{\mathcal{R}}dE \mathcal{M}\left(E\right)\mathcal{T}\left(E\right)\left(f_{L}\left(E,T_{L}\right) - f_{R}\left(E,T_{R}\right)\right). 
\label{lbfcu}     
\end{equation}
The transmission probability in Eq.~\ref{lbfcu}, for an electron to traverse the channel length (the longitudinal dimension) is $ \mathcal{T}\left(E\right) $, the function $ \mathcal{M}\left(E\right) $ is the number of modes, and $ f_{L,R}\left(E,T_{L,R}\right) $ represents the Fermi distribution at the two contacts. As a first approximation, we set the transmission to unity assuming the target structure (see Fig.~\ref{scther}) to be homogeneous throughout. To estimate $ \mathcal{M}\left(E\right) $, as is standard practice, we assume periodic boundary conditions along the \textit{y}-axis such that the $ k $-channels are equi-spaced by $ 2\pi/W $.~\cite{datta1997electronic} Here, $ W $ is the width (the transverse span along the \textit{y}-axis) of the sample. Each unique $ k $-vector is a distinct mode and the number of such momentum vectors is determined from the inequality, $ -k_{f} < k_{y} < k_{f} $. The upper and lower bounds of the inequality are the momentum vectors that correspond to the Fermi energy, E$_{f}$. The approximate number of modes is therefore $ k_{f}W/\pi $. Also, note that the integral in Eq.~\ref{lbfcu} must be evaluated over two energy-manifolds, each in the vicinity of the conduction $\left(E_{c}\right)$ and valence band $\left(E_{v}\right)$ extremum. Explicitly, the energy manifold of integration (Eq.~\ref{lbfcu}) is a simple union of two disjoint intervals given by $ \mathcal{R}_{1}: E \in \left\lbrace -\infty, E_{v}\right\rbrace $ and $ \mathcal{R}_{2}: E \in \left\lbrace E_{c}, \infty\right\rbrace $. The domain is then, $ \mathcal{R} = \mathcal{R}_{1} \cup \mathcal{R}_{2} $. Bearing these in mind, Eq.~\ref{lbfcu} for thermal current can be recast as
\begin{equation}
I_{th} = \dfrac{eW}{\pi h}\int_{\mathcal{R}}g\left(E\right)dE\int_{-\pi/2}^{\pi/2}d\theta \cos\theta \left(f_{1}\left(T_{1}\right) - f_{2}\left(T_{2}\right)\right). 
\label{lbfcun}
\end{equation}
The function $ g\left(E\right) $ in Eq.~\ref{lbfcun} is the analytic representation of the $ k $-vector in energy space. Using Eq.~\ref{bpdp} and the exchange field, the function $ g\left(E\right) $ is
\begin{equation}
g\left(E\right) = \left[\dfrac{1}{2\alpha^{2}\cos^{4}\theta}\left(\Omega - \left(2\alpha\Delta\cos^{2}\theta + \beta^{2}\sin^{2}\theta\right)\right)\right]^{0.5}  .
\label{keqn}
\end{equation} 
For brevity, we have used the short-hand notation $ \Omega = \sqrt{\left(2\alpha\Delta\cos^{2}\theta + \beta^{2}\sin^{2}\theta\right)^{2} + 4\cos^{4}\theta\left(\mathcal{E}^{2} - \Delta^{2}\right)\alpha^{2}} $ and $ \mathcal{E} = \left(E - \eta\Delta^{ex}\right) $. As a clarifying note, the sum of modes along the width $\left(W\right) $ is $\sum_{k_{f}W/\pi}\Delta k_{y} $ which in the continuum limit changes to $ \int_{-k_{f}} ^{k_{f}}dk_{y} $. As usual, the azimuthal angle is $ \theta $ while $ k_{x} = k\cos\theta $ and $ k_{y} = k\sin\theta $. Note that the limits of angular integration satisfies the span of a given $ k $-vector $\left(-k_{f} < k_{y} < k_{f} \right) $ or mode. To obtain an estimate of the thermal current, a numerical integration of Eq.~\ref{lbfcun} can be carried out. It is worthwhile to emphasize again on what we briefly alluded to above: The current in Eq.~\ref{lbfcun} has two components - from the conduction band electrons and the valence band holes - that flow in opposite directions for a certain potential drop. To further elucidate, when the left contact (in Fig.~\ref{scther}) is at a higher temperature than its right counterpart on the right, we define the electron current to flow from the left and empty in the right contact while the hole current flows in the exact opposite sense. 

A pair of noteworthy remarks is in order here: Firstly, we reiterate that the thermal current is purely an outcome of the temperature gradient induced difference in the Fermi levels of the two contacts; therefore, before any numerical value to a temperature gradient aided thermal current is sought, it must be borne in mind that in an open-circuit steady state case, a build-up of electrons on one side (due to $ I_{th}$) creates a potential drop which is countered by a oppositely-directed drift current. The net current flow is therefore zero in this arrangement. A voltmeter simply measures the voltage developed across the sample attributed to the charge separation brought about by the flow of the diffusive $ I_{th} $. For a finite current to flow under a temperature gradient, an electrical load must be hooked between the two contacts maintained at different temperatures. A qualitative description centered around this point and thermoelectric efficiency appears later in Section~\ref{tees}. \textcolor{red}{The second observation concerns the analysis presented until now which assumes that the carriers seamlessly flow from the contacts to the channel indicating a perfect transmission $\left(\mathcal{T} = 1\right)$. In practice, scattering at the boundary of the BP channel and thermal reservoirs (contacts) causes a reduction such that $ \mathcal{T} < 1 $. The scattering effects can be estimated from several approaches including the classical Boltzmann transport equation or the fully quantum mechanical non-equilibrium Green's function method. Typically, the use of metal contacts affixed to semiconducting channels creates a work function mismatch leading to an ohmic contact for \textit{p}-type sample and Schottky behaviour for its \textit{n}-doped counterpart. The Schottky barrier for the \textit{n}-channel device shows a non-linear conduction. BP transistors have been found to conform to this observation.~\cite{li2014black,perello2015high} However, for now, we ignore such complications and adopt a simple ballistic regime allowing the assignment of unity to $ \mathcal{T} $ throughout. An obvious consequence of such an assumption is an overestimation of the thermal-driven charge current.}

\begin{figure}[t!]
\includegraphics[scale=0.7]{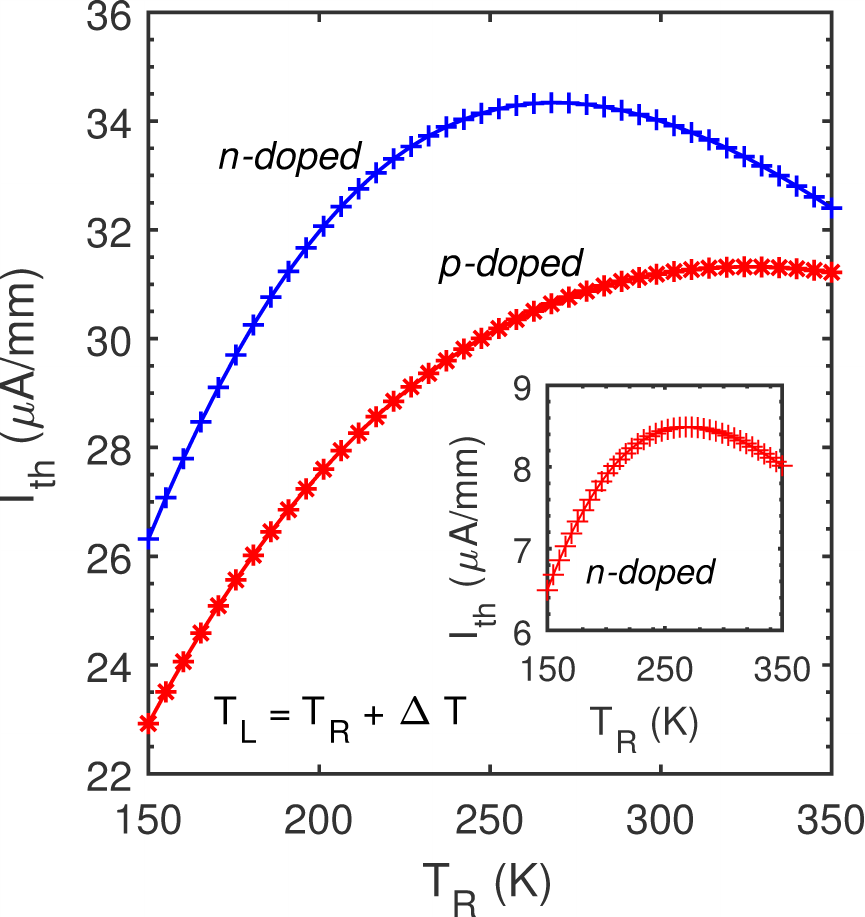}
\vspace{-0.2cm}
\caption{The numerically obtained current $\left(I\right)$ that flows in a multi-layer BP sheet clamped between two contacts at dissimilar temperatures (see Fig.~\ref{scther}) is shown. Mirroring the magnetization-induced asymmetry of the conduction and hole states, $ I $ is lower for a \textit{p}-doped material vis-\'a-vis an \textit{n}-type structure. The inset shows the current for an \textit{n}-type graphene sheet with band gap and Fermi velocity identical to the layered BP. The thermal current for gapped Dirac graphene is lower than its semi-Dirac BP counterpart. Note that the temperature $\left(T_{R}\right)$ shown along the \textit{x}-axis is for the left contact; the left contact is always set to $ T_{R} + \Delta T $. Here, $ \Delta T = 25\, K $.}
\label{thcur}
\vspace{-0.6cm}
\end{figure}

Within the framework of the discussed formalism, we can now numerically evaluate the temperature gradient driven current; to do so, we begin by fixing the parameters starting with the temperatures of the two contacts. The temperature is swept between $ T \in \left[150,350\right]\, K $ maintaining a constant difference of $ \Delta T = 25\, K $ between the left (hotter) and right contact. The exchange field of the ferromagnet that breaks the particle-hole symmetry is set to $ \Delta_{ex} = 30.0\, meV $. As for BP, from previously determined material constants, the effective mass for the parabolic branch is $ m_{eff} = 1.42m_{0} $ while the band gap with \textit{K}-doping is approximately $ \Delta = 0.36\, eV $. The BP band gap is tunable via the concentration of the \textit{K}-dopant.~\cite{baik2015emergence} The Fermi velocity for the linear branch is $ v_{f} = 5.6 \times 10^{5}\, ms^{-1}$. Note that the out-of-plane magnetization splits the conduction and valence bands into spin-up and spin-down ensembles; for the selected material constants, the bottom the conduction spin-up (down) band is $ 0.21 \left(0.15\right)\, \mathrm{eV}$. The corresponding top of the spin-up (down) valence band is $ -0.15 \left(-0.21\right)\, eV $. The Fermi distribution functions (see Eq.~\ref{lbfcun}) are computed by assigning an identical electrochemical potential $\left(\mu\right)$ to both contacts; to simulate \textit{n}- and \textit{p}-type character, we toggle $ \mu $ between $ \pm 0.17\, eV $. In each case, $ \mu $ is located between the bottom (top) of the spin-split conduction (valence) spin-up and spin-down bands. The current (see Fig.~\ref{thcur}) is obtained by a numerical integration (Eq.~\ref{lbfcun}) considering all possible modes within $ 65.0\, meV $ from the conduction and valence band extremum and a range of temperatures while holding a constant difference between the contacts. 

\textcolor{red}{The analysis presented heretofore used semi-Dirac BP to estimate thermal current under a temperature gradient. The same set of steps can be applied to undertake a comparison between the thermal currents carried by semi-Dirac and Dirac materials. A gapped Dirac material, for instance, graphene, is described in the continuum limit as
\begin{equation}
H_{Dirac} = \hbar v_{f}\left(k_{x}\tau_{x} + k_{y}\tau_{y}\right) + \Delta\tau_{z}.
\label{grham}
\end{equation}.
The symbols in Eq.~\ref{grham} have their usual meaning. Moreover, to uncover any contrasting behaviour between the semi-Dirac and Dirac cases, we assign the Fermi velocity of BP and its band gap to the graphene-like material described by Eq.~\ref{grham}. The exchange coupling as before is set to $ \Delta^{ex} = 30.0\,meV $ and the contact temperatures are varied exactly as done for BP. The thermal current for the Dirac case can be again obtained from Eq.~\ref{lbfcun} by simply changing the function $ g\left(E\right) $ to reflect the transmission modes of the graphene-like material. It is
\begin{equation}
g\left(E\right) = \dfrac{1}{\hbar v_{f}}\sqrt{\left[\left(E - \eta\Delta^{ex}\right)^{2} - \Delta^{2}\right]}.
\label{kdeqn}
\end{equation}
The two spin-states are distinguished by $ \eta = \pm 1 $. Inserting $ g\left(E\right) $ from Eq.~\ref{kdeqn} in Eq.~\ref{lbfcun} and positioning the Fermi level identically as in the case of \textit{n}-doped BP followed by a numerical integration, the $ I_{th} $ for a representative Dirac material is shown in the inset of Fig.~\ref{thcur}. The thermally-driven current is feebler vis-\'a-vis a semi-Dirac material. However, the thermal current profiles for both Dirac and semi-Dirac display a similar behaviour with a prominent plateau region in each case - a distinctive feature of currents with a purely thermal origin.}

\begin{figure}[t!]
\includegraphics[scale=0.8]{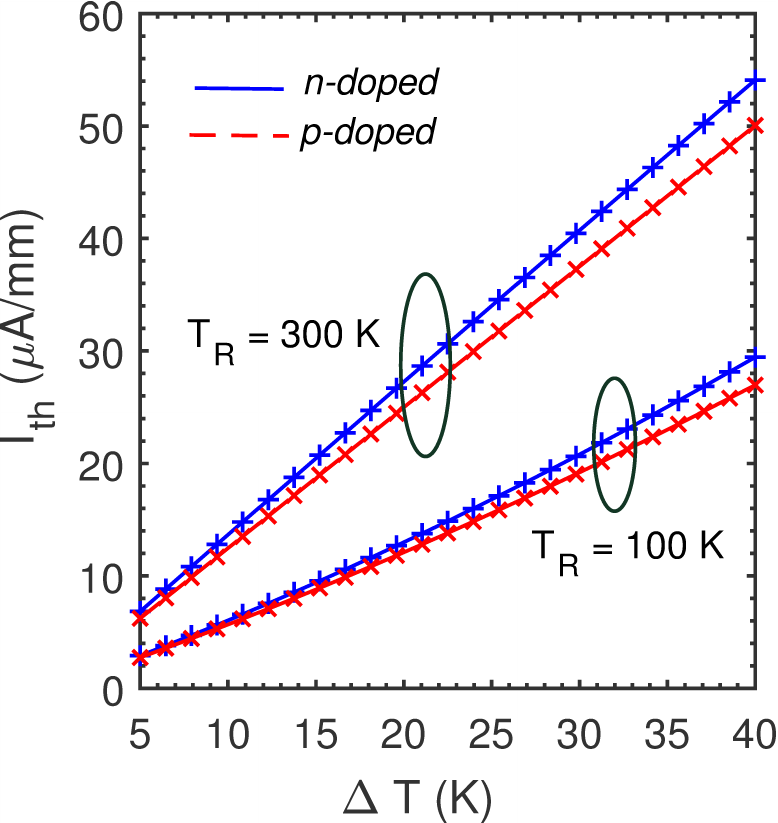} 
\vspace{-0.2cm}
\caption{The thermal current behaviour as a function of temperature difference that exists between the two contacts. The material parameters are identical to those used in Fig.~\ref{thcur}. The two temperatures $\left(T_{R} = 100, 300\, K\right)$ indicated on the plot are the starting values for the right contact, the temperature of the left contact $\left(T_{L} = T_{R} + \Delta T\right)$ is then swept up to enlarge the difference (shown on the \textit{x}-axis) with its right counterpart that progressively enhances the Fermi drop $\left(\Delta\mu\right)$ and consequently manifests as a larger thermal current.}
\label{thcurdt}
\vspace{-0.6cm}
\end{figure}

We comment on a few noteworthy features of Fig.~\ref{thcur}; firstly, it is immediately recognizable that as the temperature rises, the thermal current begins to saturate for both \textit{n}- and \textit{p}-type, a behaviour attributable to the diminishing Fermi level difference $\left(\Delta \mu\right)$ with a rise in temperature for a given energy. At hand though, there also exists a larger smearing of the Fermi function at a higher temperature opening up additional modes ($\left(M\right)$ in Eq.~\ref{lbfcun}) accessible for transport, however, in this case, the fall in $\Delta \mu$ executes the more definitive role. In the same spirit, for a fixed temperature, the current falls (see inset, Fig.~\ref{thcur}) at higher energies as $ \Delta \mu $ is again lowered for a pre-determined temperature difference between the contacts. To offer additional substantiation to this line of reasoning, a plot (Fig.~\ref{thcurdt}) of the thermal current for several temperatures differences is distinguished by an increasing behaviour, a fact that constitutes a simple demonstration of an enlarged $ \Delta \mu $. This increment to  $ \Delta \mu $ translates into a higher current. Notice that the current (for both doping cases) shown in Fig.~\ref{thcur} receives contribution from four components: They are a pair consisting of spin-up and spin-down electron components from the conduction bands and a similar but counteracting hole-based set originating in the valence bands. When BP is doped \textit{n}-type, the position of the Fermi level in the conduction band guarantees completely (almost, if smearing is accounted) filled valence bands, a consequence of which is a negligible hole current. By the same token, for a \textit{p}-type material, the conduction states are nearly empty ensuring that bulk of the current is carried by holes. From an experimental standpoint, both \textit{n}- and \textit{p}-type conduction has been observed~\cite{li2014black} in BP consistent with the narrow band gap that permits tuning of the Fermi level close to either the valence or conduction band.

\begin{figure}[t!]
\includegraphics[scale=0.78]{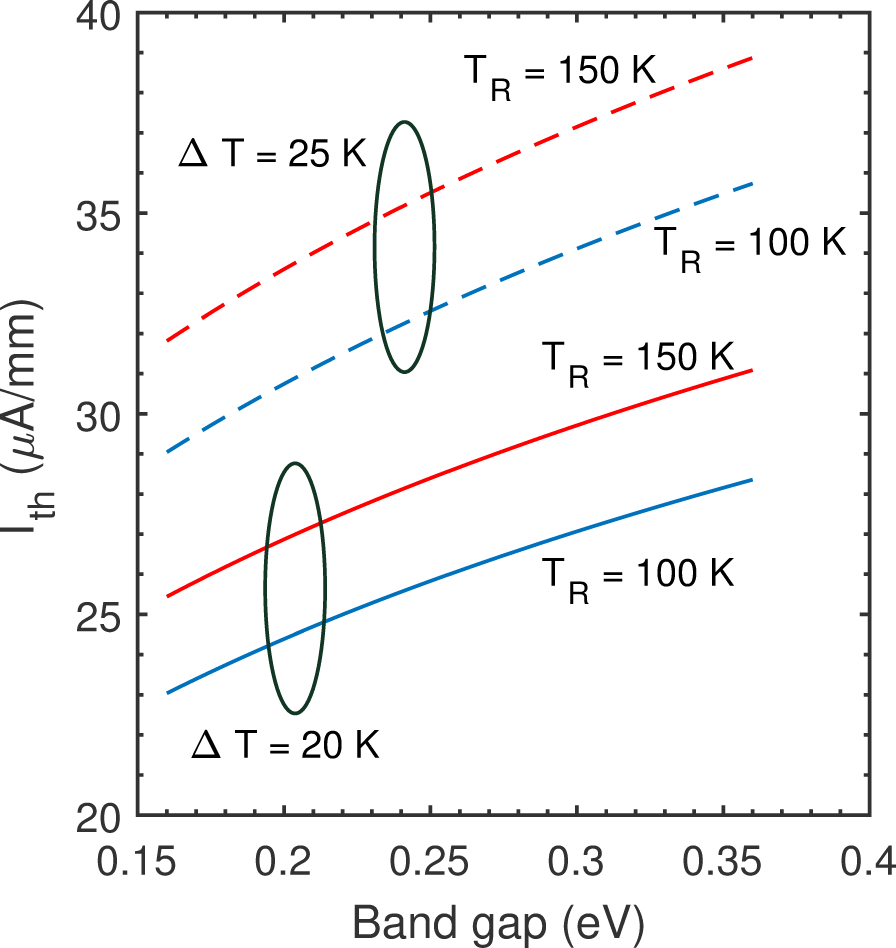} 
\vspace{-0.15cm}
\caption{The thermal current is shown for several band gaps in an \textit{n}-type BP. The current is determined for two temperature points $\left(T_{R} = 100, 300\, K\right)$ as noted in the figure. The temperature of the left and right contacts are therefore $ T_{R} + \Delta T $ and $ T_{R} $, respectively. The current curves have been shown for a pair of $ \Delta T = \lbrace 20, 25\, K\rbrace $. A higher band gap (that is capped around the $ \Delta = 0.35 \, eV $ in BP) gives rise to a larger thermal current. The thermal current was calculated taking into account conduction energy states lie within a 65.0 $\mathrm{meV} $ range from bottom of the conduction band. Note that in absence of any magnetization effects, the two spin components contribute equally to the thermal current. The corresponding case of a \textit{p}-type BP, again assuming zero magnetization and a Fermi level adjusted to top of the valence band, the thermal current is identical in magnitude to the \textit{n}-type case but with reversed polarity.}
\label{thcureg}
\vspace{-0.5cm}
\end{figure}

We pointed to the electric field aided band gap modulation of BP, either via \textit{K}-doping or an applied electric field; when such a mechanism is available, it is beneficial to check the associated variations in the thermal current. We use two temperature points $\left(T_{R} = 100, 150\, K\right)$ and for each such value, the difference, $ T_{L} - T_{R} $, is assigned a pair of numbers, $ \Delta T = 20, 25\, K $. The thermal current is plotted (Fig.~\ref{thcureg}) for several band gaps $\left(\Delta\right)$ in \textit{n}-type BP. For the purpose of illustration, the Fermi level is aligned with bottom of the conduction band (\textit{n}-type material), which is $ \Delta/2 $. The magnetization is not included and as explained above, the hole currents contribute negligibly for an \textit{n}-type BP. The thermal current increases for a larger gap, an outcome simply explained by noting that in the vicinity of a higher Fermi energy (remembering that it is aligned to the conduction band minimum) a greater number of current carrying modes are present or equivalently a higher value for $ \mathcal{M}\left(E\right) $ in Eq.~\ref{lbfcun}. The thermal current expectedly rises. For further elucidation, a more specific case can be examined that offers a more concrete feel of the magnitude of the thermal current. We consider a typical arrangement wherein the temperature is set to $ T = 300\, K $ while the difference between the contacts is maintained at a steady $ \Delta T = 25\, K $. The injected carriers are located in the vicinity of the conduction (valence) band minimum (maximum) while the magnetization has been turned off in this case. The Fermi level, as in the preceding case is adjusted to $ \mu = 0.17\, eV $, located between the mimima of the spin-split semi-Dirac BP's conduction bands. Inserting these numbers in Eq.~\ref{lbfcun}, the current approximately evaluates to $ I_{th} \approx 31 \mu A/mm $. A caveat about this result and in general pertaining to this work must however be noted: First and foremost, the current solely driven by a temperature difference is the maximum achievable for the selected thermal settings; the quantum of current flow though in a realistic setup will degrade primarily through electron-phonon scattering which is active in multi-layer BP for $ 100\, K < T <  300\, K $. The impurity scattering dominates below the $ T = 100\, K $ mark. For a comparative account of current flow, it is useful to quote measurement data collected by Liu \etal in field-effect transistor (FET) structures with stacked BP (stacked BP can be thought of as a bulk form) wherein they report~\cite{liu2015semiconducting} a maximum drain current of $ I_{DS} = 194\,mA/mm $ in an FET structure at drain bias $\left(V_{DS}\right)$ of 2.0\,V and $ 1.0\,\mu m $ in channel length. This current under an electric bias is roughly three orders larger over $ I_{th} $. In passing, it is pertinent to mention that a converse of this thermionic process - the electrocaloric response - exists with similar applications in solid state refrigeration and Peltier cooling. The electrocaloric effect centers around a reversible change in temperature by an electric signal and has been demonstrated~\cite{moya2014caloric,mischenko2006giant} in ferroelectrics such as PbTiO$_{3}$. We do not discuss it here. 

\vspace{0.3cm}
\section{Electronic part of the heat capacity}
\vspace{0.3cm}
The preceding analysis focused on an elementary Peltier cooling~\cite{grosse2011nanoscale} model wherein a current flows under a temperature gradient removing the surplus heat. However, certain applications require the ability to maintain temperatures, an attribute tied to a large heat capacity. We examine the heat capacity of semi-Dirac BP employing the Sommerfeld expansion (SE). The SE retaining up to only the second order term is~\cite{grosso2014solid}
\begin{equation}
\int_{-\infty}^{\infty}H\left(\epsilon\right)f\left(\epsilon\right)d\epsilon = \int_{-\infty}^{\mu}H\left(\epsilon\right)d\epsilon + \dfrac{\pi^{2}}{6}\left(k_{B}T\right)^{2}H^{'}\left(\mu\right).
\label{bse}
\end{equation}
Briefly, $\epsilon $ denotes energy, $ k_{B} $ is the Boltzmann constant, and the function $ H\left(\epsilon\right) $ is expanded in the vicinity of $ \mu $, the Fermi level. For a heat capacity derivation, we start by setting the function $ H\left(\epsilon\right) = \epsilon\mathcal{D}\left(\epsilon\right) $ and inserting in Eq.~\ref{bse} gives the internal energy $ U = \int \epsilon\mathcal{D}\left(\epsilon\right)d\epsilon + \left(\pi k_{B}T\right)^{2}\mathcal{D}\left(\epsilon\right)/6 $. The density of states (DOS) is $ \mathcal{D}\left(\epsilon\right)$. Differentiating the internal energy expression w.r.t T, we get the heat capacity at constant volume. It is given as 
\begin{equation}
C_{v} = \dfrac{\partial U}{\partial T} = \dfrac{\pi^{2}k_{B}^{2}T}{3}\mathcal{D}\left(\epsilon\right).
\label{sphc}
\end{equation}

A numerical value for the heat capacity therefore entails an evaluation of $\mathcal{D}\left(\epsilon\right)$. For DOS calculation of anisotropic BP, we use the dispersion in Eq.~\ref{bpdp} to write
\begin{equation}
\begin{aligned}
\mathcal{D}(\epsilon) &= \int \dfrac{d^{2}k}{\left(2\pi\right)^{2}}\delta\left(\epsilon - E(k)\right), \\
& = \dfrac{1}{\left(2\pi\right)^{2}}\sum\limits_{j}\int_{0}^{2\pi} d\theta \dfrac{k_{j}}{\vert g^\prime(k_{j})\vert}.
\label{doseq}
\end{aligned}
\end{equation}
In Eq.~\ref{doseq}, the azimuthal angle is $ \theta $. We have used the identity $ \delta\left(g(x)\right) = \sum\limits_{j}\dfrac{\delta(x-x_{j})}{\vert g^\prime(x_{j})\vert}$ such that $ g(x_{j}) = 0 $ and $ x_{j} $ is a simple zero of $ g(x) $. The function $ g(k) $ in this $ \epsilon - \sqrt{\left(\Delta + \alpha k^{2}\cos^{2}\theta\right)^{2} + \left(\beta k\sin\theta\right)^{2}} $. Setting $ g(k) = 0 $ and solving, the positive root $\kappa $ can be expressed as $\kappa^{2} = \left(\sqrt{p^{2} + 4\alpha^{2}\cos^{4}\theta\left(\epsilon^{2} - \Delta^{2}\right)} - p\right)/\left( 2\alpha^{2}\cos^{4}\theta\right) $, where $ p = \left(2\Delta \alpha\cos^{2}\theta + \beta^{2}\sin^{2}\theta\right) $. Inserting the positive root and the derivative (w.r.t $ k $ at $ \kappa $) of $ g\left(k\right)$, the expression for DOS is
\begin{equation}
\mathcal{D}\left(\epsilon\right) = \dfrac{1}{4\pi^{2}}\int_{0}^{2\pi} d\theta\dfrac{E\left(\kappa\right)}{2\alpha\cos^{2}\theta\left(\Delta + \alpha\kappa^{2}\cos^{2}\theta\right) + \left(\beta\sin\theta\right)^{2}}.
\label{dosfe} 
\end{equation}
The DOS for semi-Dirac BP when substituted in Eq.~\ref{sphc} gives the heat capacity at constant volume. This calculation is not generalized and is true only when a Sommerfeld expansion $\left(T\ll T_{f}\right)$ is possible. The Fermi temperature, $ T_{f} = \epsilon_{f}/k_{B} $, $ \epsilon_{f} $ denotes the Fermi energy, and $ k_{B} $ is the Boltzmann constant. The crux of a $ C_{V} $ calculation is then a determination of DOS; to that end, we perform a numerical integration of the DOS integral (Eq.~\ref{dosfe}) for a range of energies in the vicinity of the conduction band minimum. The numerically calculated DOS multiplied with the appropriate pre-factor in Eq.~\ref{sphc} furnishes the $ C_{V} $ for a two-dimensional semi-Dirac electron BP. In this calculation (see Fig.~\ref{sphc}), the BP sample area was set to $ A = 1\, cm^{2} $. The heat capacity within the framework of the Sommerfeld expansion is simply a portrayal of the DOS making it amenable to changes via processes and excitations that can renormalize the armchair mass or alter the zigzag Fermi velocity. 
\begin{figure}[b!]
\includegraphics[scale=0.75]{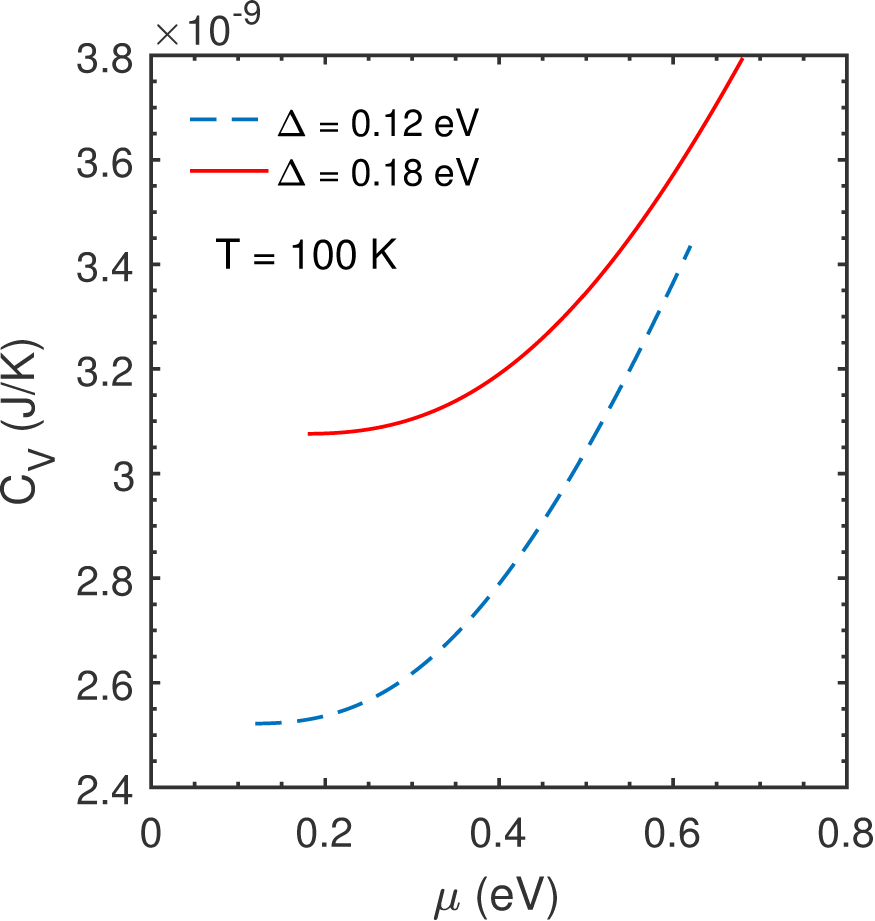} 
\vspace{-0.05cm}
\caption{The electronic contribution to the heat capacity $ \left(C_{V}\right)$ of the two-dimensional semi-Dirac electron BP gas is plotted for several Fermi energies and two band gaps. The temperature for this calculation was set to $ T = 100\, K $ and the area of the sample is assumed to be $ A = 1\, cm^{2}$. For the range of Fermi energies assumed, the Fermi temperature $\left(T_{f}\right)$ is much larger than $ T = 100\, K $. Note that in this regime, the heat capacity is of the form $ C_{V} = \gamma\, T $, where $ \gamma $ is the so-called Sommerfeld coefficient.}
\label{hcp}
\vspace{-0.3cm}
\end{figure}

At this stage, it is somewhat instructive to compare the heat capacity of semi-Dirac BP with an identically-sized sample of two-dimensional electron gas (2DEG) with conventional parabolic dispersion. For ease of comparison, we select conduction electrons of zinc blende GaAs at $ T = 100\, K $ with an effective mass $ m_{e} = 0.067m_{0}$ and a direct band gap of $ E_{g} = 1.41\, eV $ at the $ \Gamma $ point of the Brillouin zone. The  heat capacity in this case (using Eq.~\ref{sphc}) is $ C_{V} = A\pi m_{e}k_{B}^{2}T/(3\hbar^{2}) $. In writing this 2DEG $ C_{V} $ formula, we used the 2D DOS expression of $ m_{e}/\pi\hbar^2 $. Plugging in the appropriate material parameters and setting $ A = 1\,cm^{2} $, we have $ C_{V} = 1.1 \times 10^{-11}\, J/K $. As a simple check to determine the applicability of the Sommerfeld expansion for calculating the heat capacity of GaAs, we set the electron density to $ n = 10^{11}\, cm^{-2} $ which corresponds to a Fermi wave vector of $ k_{F} \approx 0.008\, A^{-1} $. The Fermi energy is therefore (from the bottom of the conduction band which is at $ E_{g} $) $ \epsilon_{f} = 3.6\, meV $ and the equivalent Fermi temperature is approximately $\left(T_{f} = 1.6 \times 10^{4}\, K\right)$. Since $ T_{f} \gg T $, the use of an SE-derived $ C_{V}$ formula is correct. To compare, the heat capacity for the model GaAs 2DEG is roughly three-orders reduced than a representative value (see Fig.~\ref{hcp}) of $ 3 \times 10^{-9} J/K $ for semi-Dirac hybrid BP. It is of some consequence to observe that the low $ C_{V} $ values are in part attributable to the shrunken-dimensionality (2D) of the system (compared to the three-dimensional bulk) and further accentuated in the case of GaAs due to the intrinsic smallness of effective mass $\left(0.067m_{0}\right)$ of conduction electrons around the Brillouin zone centre. In sharp contrast, the corresponding value for \textit{K}-doped BP at $ 1.42m_{0} $ along the parabolic branch is roughly twenty-fold higher. Additionally, while the $ C_{V} $ for GaAs conduction electrons confined in a quantum well is constant around the Brillouin zone centre (assuming the effective mass does not vary too much), the hybrid BP shows a rise attributable to the linear branch of the dispersion along the zigzag axis. The contribution of this branch to the DOS rises linearly with energy; however, its complete manifestation is suppressed by the parabolic branch forbidding a truly graphene-like behaviour. For small momentum values close to the Dirac crossing, the linear branch is dominant and the DOS rises with energy marked by an accompanying increase in the heat capacity. 

In light of the demonstrated higher $ C_{V} $ of semi-Dirac BP, it is reasonable to anticipate that a system with such hybrid dispersion can serve as a better coolant to hold the temperature steady for a longer period compared to a 2D nanostructure with parabolic materials. It therefore also follows that techniques to augment the linear part of the dispersion over the parabolic branch can offer greater advantage, especially when the DOS is inextricably linked in the unfolding of a physical phenomenon. A few notable examples in this regard could be quantum capacitance and paramagnetic susceptibility, both quantities are linearly connected to the DOS. We briefly discuss novel techniques to modulate the DOS in the closing section. 

\vspace{0.3cm}
\section{Thermoelectric Efficiency}
\label{tees}
\begin{figure}[t!]
\includegraphics[scale=0.65]{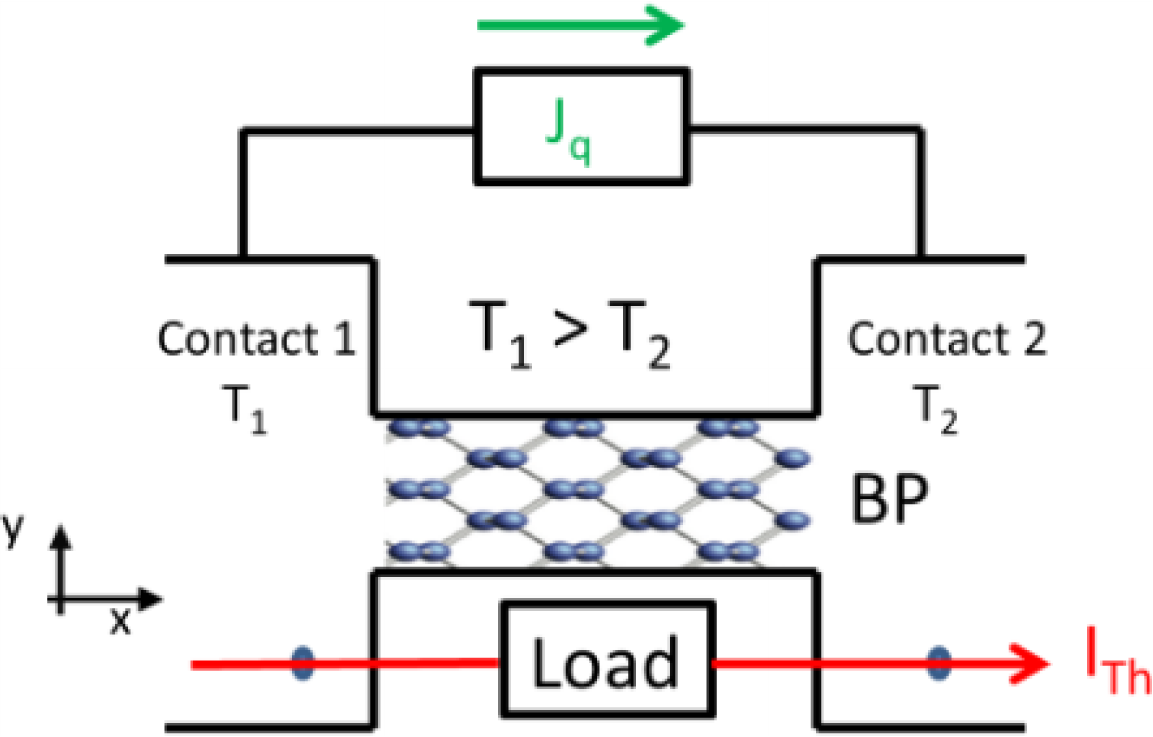} 
\vspace{-0.05cm}
\caption{The schematic of an energy generator that `outputs' a thermoelectric current $\left(I_{th}\right)$ as a response to an `input' heat current, $ J_{q} $. The temperature gradient between the contacts is maintained by the heat current. The ratio $ I_{th}/J_{q} $ is the Carnot-like thermoelectric efficiency of this setup.}
\label{teesch}
\vspace{-0.3cm}
\end{figure}

\vspace{0.3cm}
The thermal current and heat capacity calculations, besides their apparently `independent' roles in power conversion and temperature control, can be brought together to compute a Carnot-like thermoelectric efficiency (TEE) under a temperature gradient. While we have considered the case where the temperature gradient is a natural outcome of the microscopic processes in a miniaturized device, it is possible to recreate a similar arrangement by pumping an external heat current $\left(J_{q}\right)$ between the contacts. The resulting temperature difference drives an electric current $\left(J_{e}\right)$, a process to convert heat into useful work. A TEE for the operation of this prototypical thermoelectric engine (for a schematic, see Fig.~\ref{teesch}) can be defined as $ \eta = I_{th}/J_{q} $. In the following, we derive an expression for TEE that relevant to the energy cycle illustrated in Fig.~\ref{teesch}.  
To begin, first note that for setting up a certain temperature gradient, $ \Delta T $, the desired heat current from Fourier's law is $ J_{q} = -\kappa_{q}\Delta T $. Here, $\kappa_{q}$ is the complete thermal conductivity and is the sum of the electronic $\left(\kappa_{el}\right)$ and lattice $\left(\kappa_{el}\right)$ contributions. We do not analyze the lattice contribution, simply noting that it can be established from the kinetic theory using the relation~\cite{kim2016electronic} 
\begin{subequations}
\begin{equation}
\kappa_{ph} = \dfrac{1}{3}C_{v,ph}v_{ph}\Lambda_{ph}.
\label{kapke}
\end{equation}
The phonon velocity is $ v_{ph} $ while $ \Lambda_{ph} $ indicates the corresponding mean free path. The specific heat for the lattice part $\left(C_{v,ph}\right)$ can be expressed via the Debye model as~\cite{dove1993introduction}
\begin{equation}
C_{v,ph} = \dfrac{9Nk_{B}T^{3}}{\Theta_{D}^{3}}\int_{0}^{\Theta_{D}/T}\dfrac{x^4e^{x}}{\left(e^{x}- 1\right)^{2}}dx.
\label{debyeph}
\end{equation}
\end{subequations}
In Eq.~\ref{debyeph}, $ N $ is the number of vibrational modes with frequency $ \omega $, the Boltzmann constant is $ k_{B} $, $ \Theta_{D} $ is the Debye temperature, and $ x = \hbar\omega/k_{B}T $. For the electronic part, a similar kinetic theory description can be written, replacing $ C_{v,ph} $ with $ C_{v, el} $ in Eq.~\ref{debyeph}, $ v_{ph} $ and $ \Lambda_{ph} $ with the Fermi velocity and the electronic mean free path, respectively. Notice that the electronic contribution to $ C_{V} $ can be obtained making use of Eq.~\ref{sphc} for semi-Dirac BP, the material of interest in this work. The input power, which is the pumped heat current, by clubbing the two (electronic and lattice) thermal conductivity contributions (Eqs.~\ref{sphc} and ~\ref{debyeph}) and inserting in Fourier's heat law gives
\begin{equation}
J_{q} = \dfrac{1}{3}\left(C_{v,ph}v_{ph}\Lambda_{ph} + C_{v,el}v_{f}\Lambda_{el}\right)\Delta T.
\label{fhq}
\end{equation}
The output power is $ I_{th}^{2}R $, where $ I_{th} $ can be determined from Eq.~\ref{lbfcun} and $ R $ is the resistive load. The Carnot-like TEE is therefore
\begin{equation}
\eta = \dfrac{3J_{e}^{2}R}{\left(C_{v,ph}v_{ph}\Lambda_{ph} + C_{v,el}v_{f}\Lambda_{el}\right)\Delta T}.
\label{tee}
\end{equation}
A more careful analysis of thermoelectric efficiency must however start from the coupled energy (heat) and charge equations~\cite{duan2016high} and account for the overall Seebeck-induced current that may flow in a complete circuit, which could be quantitatively different from the thermionic contribution predicted by $ I_{th} $ of Eq.~\ref{lbfcun}. The coupled heat and electric current equations are
\begin{equation}
\begin{aligned}
J_{e} = L_{11}E + L_{12}\left(-\Delta T\right), \\
J_{q} = L_{21}E + L_{22}\left(-\Delta T\right).
\label{eqen}
\end{aligned}
\end{equation}
From Eq.~\ref{eqen}, the Seebeck coefficient is $ S $ given by $ L_{12}/L_{11} $ while the electric (thermal) conductivity is simply $ L^{11}\left(L^{22}\right)$. The thermoelectric current $\left(J_{e}\right)$ in absence of any external bias $\left(\mathbf{E}= 0\right)$ has an additional contribution $ \sigma S \Delta T $ which must be added to the total output thermal current thus modifying the TEE expression in Eq.~\ref{tee}. The complete set of derivations accounting for these processes that amends TEE is postponed to a subsequent work.

\vspace{0.3cm}
\section{Closing Remarks}
\vspace{0.3cm}
We have presented theoretical calculations on thermal current that flows under an impressed temperature gradient in a two-dimensional semi-Dirac system. The layered black phosphorus, for reasons rooted in its innate anisotropy (briefly described in Section.~\ref{sec1}) was the chosen representative semi-Dirac material. The heat capacity of BP was also computed in the regime where the Sommerfled expansion remains valid. The flow of a thermal current and the attendant heat removal represents a possible realization of the Peltier cooling while the heat capacity, on the contrary, signifies the intrinsic effectiveness of absorbing heat and maintaining a steady temperature. These seemingly distinct phenomena, however, in a low temperature regime (that ignores phonon contributions) show a shared association at the microscopic level. We elaborate on this by recalling that while the Landauer formalism~\cite{stone1988measured} works with the density of modes, essentially an enumeration of the probable momentum pathways for carriers to flow, the heat capacity explicitly uses the density of states. A dominant role for either heat removal or absorption is therefore a function of the density of states $ \left(\mathcal{D}\left(\epsilon\right)\right) $. A slightly different perspective can also be offered in way of connecting the $ \mathcal{D}\left(\epsilon\right) $ to the temperature-driven current through the intrinsic Seebeck coefficient $\left(\mathcal{Q}_{ij}\right)$ of the material. Observe that the thermovoltage $\left(V_{L,hot} - V_{R,cold}\right)$ developed between the two ends of the material can be indirectly gauged via the Seebeck coefficient, usually expressed in units of $ \mu V/K $. We define the functional form of $ \mathcal{Q}_{ij} $ using the well-known Mott relationship.~\cite{jonson1980mott} In tensorial form, where $\hat{i}$ and $ \hat{j}$ are unit vectors in the 2D plane, it is given by 
\begin{equation}
\mathcal{Q}_{ij} = -\dfrac{\pi^{2}}{3e}\dfrac{k_{B}^{2}T}{\sigma}\dfrac{\partial \sigma_{ij}}{\partial \epsilon}.
\label{mott}
\end{equation}
The Mott relation is electric conductivity $\left(\sigma\right)$ dependent, which (for an isotropic case) in turn can be written as~\cite{adam2009theory} 
\begin{equation}
\sigma = \dfrac{e^{2}v_{f}^{2}}{2}\int \mathcal{D}\left(\epsilon\right)\tau\left(\epsilon\right)\left(\dfrac{-\partial f}{\partial\epsilon}\right)d\epsilon.
\label{condexp}
\end{equation}
The $ \mathcal{D}\left(\epsilon\right)$ reappears in Eq.~\ref{condexp}, from which we conclude that the thermal current and heat capacity are bound at the microscopic level through the arrangement of electronic states. In general, the thermopower is a tensor quantity with off-diagonal elements when set out in matrix form. It can assume a diagonal-only representation for a truly isotropic (and zero magnetic field) system such as pristine graphene, unlike semi-Dirac BP. 

A suitable modification of the $ \mathcal{D}\left(\epsilon\right)$ may therefore, setting aside considerations of thermal conductivity, serve as an effective tool to tune the thermal current and heat capacity. In context of semi-Dirac BP, it was shown in a recently published work~\cite{sengupta2017anisotropy} the possibility of enhancing the $\mathcal{D}\left(\epsilon\right)$ using intense illumination. Succinctly, the dispersion of irradiated BP is revamped through the introduction of a frequency $\left(\omega\right)$ dependent linear term of the form $ \mathcal{F}\left(\omega\right)k_{x} $. The photo-induced $ \mathcal{F}\left(\omega\right)k_{x} $ in conjunction with the intrinsic linear term in the Hamiltonian (Eq.~\ref{ham1}) for low values of momentum around the semi-Dirac crossing dominates the quadratic component imparting a quasi-linear character. Since for materials with a linear dispersion the $ \mathcal{D}\left(\epsilon\right)$ monotonically increases with energy, a similar behaviour may be expected for irradiated-BP. The modulation of DOS in the quasi-linear case evidently centers around the strength of the linear terms $ \beta = \hbar v_{f} $ and $ \mathcal{F}\left(\omega\right) $ and their combined outweighing of the quadratic $ k_{x}^{2} $ armchair dispersion. 

Briefly, it is relevant to state here that BP which turns semi-Dirac with \textit{K}-doping can undergo a topological phase transition with increasing dopant concentration through a vanishing of the band gap. Beyond the band gap ceasing, for a higher \textit{K}-dopant concentration, paves the way for band inversion, a precursor to topologically protected states; indeed, such states with massless and anisotropic Dirac fermions have been observed in BP.~\cite{baik2015emergence} A simplified form of the Hamiltonian for the topological insulator (TI) can be written as
\begin{equation}
\mathcal{H}_{TI} = \hbar\,v_{x}\left(k_{x}-k_{D}\right)\sigma_{x} + \hbar\,v_{y}k_{y}\sigma_{y},
\label{ham5} 
\end{equation}
where $ v_{x} $ and $ v_{y} $ are the velocities along the \textit{x}- and \textit{y}-axes, respectively at the Dirac point. A Dirac graphene-like dispersion (Eq.~\ref{ham5})is characterized by a DOS that linearly scales with $ \vert \epsilon \vert $ and may therefore exhibit (for a finite energy) a larger heat capacity in comparison to a semi-Dirac or parabolic 2D nanostructure. It is decidedly an attractive proposition to observe topological phase transitions via doping and the concurrent advatanges tht accrue; however, the benefits of the strong anisotropy of BP, particularly suited for a large $ ZT $ may disappear. A careful set of experimental measurements may uncover the precise connection between topological phase transitions and the overall thermoelectric behaviour in semi-Dirac BP.

\vspace{0.3cm}
\section{Summary}
\vspace{0.3cm}
In closing, we have through analytic calculations predicted the thermal current and heat capacity of BP with hybrid semi-Dirac structure. Further refinements to the calculations can be carried out through inclusion of scattering in the channels suggesting that the assumed transmission is then no longer unity; besides, the role of two contacts maintained at different temperatures in a more accurate formulation must be accounted, for example, using the non-equilibrium Keldysh formalism.~\cite{jishi2013feynman} The Keldysh formalism, apart from a more accurate modeling of contacts, is also necessary as demonstrated in a recent work~\cite{eich2016temperature} that analyzed flow of charge and heat current in a nanoscale conductor taking into account transient particle behaviour that develops from an abrupt application of a temperature gradient. This work ignores any non-equilibrium heat and current density. Further, these calculations can also be applied to any two-dimensional system, a prominent example of it being the honeycomb lattice of a single layer transition metal dichalcogenide (TMDC). A TMDC single layer is naturally in accord with the requirement of particle-hole asymmetry for a finite thermal current, fulfilled via the strong spin-orbit coupling operational in the valence bands~\cite{xiao2012coupled,2053-1583-2-2-022001} and its much weaker and inequivalent counterpart in the conduction states.~\cite{wang2015physical} The inequality of the two spin-orbit couplings asymmetrically splits the gapped single layer TMDC energy levels around the high-symmetry $ K $ and $ K^{'} $ valley edges. A magnetic field which may be required in case of \textit{K}-doped BP with zero band gap to create asymmetry is therefore not a necessary condition. 

%\bibliographystyle{apsrev}
%\bibliography{reference}

\end{document}